\pdfoutput=1
\documentclass{PoS}

\usepackage{epsfig,times,color,cite,adjustbox,caption,wrapfig,lipsum,booktabs,comment,float}

\title{A Monte Carlo study of the relevance of fluorescence radiation in VHE gamma ray observations with Cherenkov telescopes}

\ShortTitle{Fluorescence}

\author{
D. Morcuende$^{a}$, J.L. Contreras$^{a}$, J. Rosado$^{a}$, \speaker{F. Arqueros}$^{,\;a}$, L. Saha$^{a}$, V. de Souza$^{b}$\\
\llap{$^a$}Departamento de F\'{i}sica At\'{o}mica, Molecular y Nuclear and UPARCOS\\
Universidad Complutense de Madrid, 28040 Madrid, Spain\\
\llap{$^b$}Instituto de F\'{i}sica de S\~{a}o Carlos\\
Universidade de S\~{a}o Paulo, 13566-590 S\~{a}o Paulo, Brazil\\
E-mail: \email{dmorcuen@ucm.es}
}

\abstract{It is generally assumed that fluorescence radiation does not play a significant role in the performance of Cherenkov telescopes. However, this assumption is required to be verified using detailed Monte Carlo simulations. In order to do this, we have implemented the production and tracking of fluorescence radiation inside the CORSIKA code, and simulated gamma-ray induced showers in the very high energy range. The most accurate fluorescence-yield data available so far was used for this purpose.

The distribution of both light components on the ground has been studied as a function of various parameters affecting the detection and reconstruction of gamma-ray showers such as the angular aperture. From these distributions, we determined the conditions under which fluorescence radiation becomes significant. These results will also be useful to estimate the corresponding systematic errors in Cherenkov telescope observations.

The full simulation results have been cross-checked, on a small sample of events, against numerical calculations based on a one-dimension shower profile and found to be compatible with each other. Both tools can be used for further investigations, like studying the possibility to modify Cherenkov telescopes for the measurement of fluorescence induced by extensive air showers.
}

\FullConference{35th International Cosmic Ray Conference - ICRC217-\\
		10-20 July, 2017\\
		Bexco, Busan, Korea}

\begin{document}

\section{Introduction}
\label{sec:introduction}

 The emission of fast light pulses is one of the hallmarks of Extensive Air Showers (EASs) arising from the interaction of very high energy cosmic particles with the Earth's atmosphere. As the light produced is spread over huge areas and a large fraction of it reaches the ground, this provides a mean to efficiently detect EASs, achieving reasonably large effective areas. Light emitted by an EAS is due to two main processes: Cherenkov effect and fluorescence emission.
 
  Charged particles traveling faster than light in the medium ($\beta > 1/n$) induce the emission of Cherenkov radiation. It forms a cone around the particle direction whose opening angle is determined by the condition $\cos{\theta}=\frac{1}{\beta n}$. As EASs contain many highly relativistic particles, the total Cherenkov radiation is peaked at a small angle of $\sim 1^{\circ}$ around the shower axis. The intersection of this angle with ground defines the so called {\it Cherenkov light pool}, beyond which the light density decreases very fast. The fact that light and particles travel at similar speeds and in roughly the same direction causes a beaming effect that compresses the light from the whole shower development in a pulse of a few nanoseconds length. The frequency spectrum of the Cherenkov light is flat in the optical range.
  
  The detection of atmospheric Cherenkov radiation from EASs is the basis of several observational techniques, being Imaging Atmospheric Cherenkov Telescopes (IACTs) the most successful one. Large telescopes equipped with cameras that record the fast Cherenkov pulses image the shower stereoscopically from different positions. IACTs have fields of view of a few degrees and point at the direction of the shower axis. They work in a typical spectral window of $300 - 500$~nm to minimize the effect of the night sky background. The best example of the next-generation telescope system where this technique will be used is the Cherenkov Telescope Array (CTA), which is currently entering into its construction phase. CTA is designed to cover the widest energy range a Cherenkov observatory has ever aimed at, from 20 GeV to 300 TeV \cite{cta-performance}.
  
  In an EAS, fluorescence light is produced by the de-excitation of atmospheric molecules previously excited by charged particles of the shower \cite{NJP}. In the wavelength interval of $300 - 450$~nm, atmospheric fluorescence is mainly produced by N$_2$ molecules with a yield of about 20 photons per MeV deposited in the atmosphere. It is important to note the strong overlapping of the fluorescence spectral interval with that of the Cherenkov telescopes. Due to the isotropic nature of the fluorescence emission, the spatial distribution of light at ground decreases gently with the distance to the shower core, in contrast with the well-defined Cherenkov light pool. Although the radiative lifetimes of involved N$_2$ states are of some tens of nanoseconds, the emission in the lower atmosphere is shortened down to a few nanoseconds due to collisional quenching. 
  
  Fluorescence telescopes have been used since many years to measure the longitudinal development of EASs in the Ultra High Energy (UHE) range ($>10^{17}$~eV). These telescopes have a wide field of view and register the  track of the shower in the atmosphere within a typical time window of microseconds. The fluorescence technique is being successfully employed in the {\it Pierre Auger Observatory} \cite{Auger}.

\subsection{Motivation}

  In the field of Cherenkov Astronomy, it is generally assumed that the contribution of the fluorescence light to the telescopes signals is negligibly small in comparison with the Cherenkov one. The reasoning is twofold: fluorescence production in EASs is less efficient (e.g., almost a factor of ten lower for 1~GeV electrons near ground), and it is spread over a large area due to its isotropic nature. However, as far as we know, the contamination of the fluorescence light in IACTs has not been quantified yet. Even though it is small, it is a systematic effect, consistently increasing the amount of light detected with respect to that expected without this contribution. The correction will become increasingly noticeable as the precision of the telescopes improves.  
  
  As a consequence of this assumption, fluorescence production is not considered in the simulation codes used to design IACTs or compute their instrument response functions. For instance, the Monte Carlo code CORSIKA \cite{corsika}, which is widely used for simulating the development of EASs to understand the performance of Cherenkov detectors, keeps in detail records of all secondary particles including the Cherenkov photons, but neglects fluorescence emission.
  
  In the field of UHE cosmic ray detection using the fluorescence technique, although fluorescence light dominates the signal, the Cherenkov component must also be included in the analysis. In fact, it is very relevant at some shower orientations. Simulations of both fluorescence and Cherenkov light are customarily based on the generation of photons from a one-dimension shower profile either obtained from an analytical EAS model or produced by CORSIKA \cite{FDSim}. Although this procedure is sufficient for most of the analyses of UHE cosmic ray showers, it neglects the lateral distribution of EASs and some other features that would be relevant for IACTs.
  
  In a different approach, V. de Souza et al.~\cite{Souza} implemented an algorithm for the fluorescence emission in the CORSIKA framework, allowing the simulation of this component with the same detail as the Cherenkov one. The code, however, was only tested for UHE cosmic rays with fluorescence telescopes. In the present work, we have followed a similar approach by incorporating the fluorescence emission in the CORSIKA code with the most accurate fluorescence data available so far (see section~\ref{sec:Tools}). Calculations based on a one-dimension shower profile (section~\ref{ssec:Calculations}) validated the simulations. As described in section~\ref{sec:Results}, this procedure has allowed us to quantify the fluorescence contamination in IACT signals for a number of cases.
  
  The simulation approach described here will also be useful to study proposals to operate IACTs in {\it Fluorescence mode} \cite{ICRC2015}, which could use the existing hardware, and other detector concepts of Cherenkov-Fluorescence telescopes \cite{Ashra,Taiga,Lhaaso}. In addition, the increasing availability of computer power makes it thinkable to include the full implementation of the fluorescence component in the IACT simulation codes even if it is a second order effect.

\section{Tools}
\label{sec:Tools}

  Two different tools have been developed in this work. Firstly the production of fluorescence light by the shower particles has been implemented in the CORSIKA code, version 7.5600. The second tool is a numerical model based on a simplified one-dimension shower profile to check the Monte Carlo simulations and that also allows for an easier interpretation of the results. 

\subsection{Implementation of fluorescence emission in CORSIKA}
\label{ssec:Insertion}

The general strategy followed for the simulation of fluorescence light in CORSIKA is very similar to the one used for Cherenkov radiation. In both cases, photons are generated each time a charged particle is propagated between two interaction points. The ionization energy loss in the step is used to evaluate the fluorescence intensity. Bunches of photons with common time and spatial coordinates, propagation direction and wavelength are produced instead of single photons. This procedure reduces significantly the computation and storage needs. Possible artificial fluctuations of this method are controlled by predefining the maximum bunch size for Cherenkov and fluorescence light separately.

We use the CORSIKA option for grids of Cherenkov detectors at the ground. Only those Cherenkov photons impinging a predefined {\it detector} area at ground are registered. The same condition was imposed to fluorescence photons. 

The workflow of the implemented simulation is as follows. The main fluorescence routine is called passing as input parameters the type of particle along with the initial and final positions, times and energies of the particle step. Electrons and positrons, which are the dominant charged particles in the shower, have small step paths (typically $<20$ m). So the number of fluorescence photons to be emitted is calculated from the energy deposited by the particle in the whole step but taking the atmospheric characteristics at the mid point. Then the step is subdivided into smaller intervals of equal length such that the ratio between the number of fluorescence photons of the step and the number of sub-steps is less than the predefined bunch size. The algorithm produces as many photon bunches of equal size as sub-steps, with the emission point and time of each bunch corresponding to the halfway sub-step. A random direction of propagation is assigned to the bunch according to an isotropic distribution, but restricted to downward directions to save computing time. The wavelength is randomly generated from the fluorescence spectrum calculated for the atmospheric characteristics at the mid point of the step. Finally, the photon bunch is transported down to observation level and the relevant data, e.g., position, arrival time, direction and wavelength, are written to a file.

For charged particles having long mean free paths (e.g., muons), the transportation step is first broken in steps of 20~m or less, each one dealt with as a separate step in the above paragraph. This is done to take into account variations of the atmosphere characteristic along the particle path. The energy deposit of the original long step is assumed to be distributed proportionally to the thicknesses of the new ones.

When the kinetic energy of a particle falls below a certain energy, the particle is no longer transported in CORSIKA and the remaining kinetic energy is deposited at the end of the step. This feature does not affect the simulation of the Cherenkov radiation provided that this energy cut is lower than the energy threshold for Cherenkov production. On the other hand, this energy deposit must be counted for fluorescence production. Therefore, it is passed to the main fluorescence routine assuming a zero length step. 

In a second step, the information stored in the output file is processed by a set of dedicated python scripts to obtain the density distribution of Cherenkov and fluorescence photons on the ground. Restrictions to the direction of propagation of bunches can be applied to include only those within the field of view of a hypothetical grid of telescopes filling the predefined detector area and pointing to a certain sky position.

The fluorescence production upon the passage of an energetic charged particle in the air is assumed to be independent of the type of particle and proportional to the energy deposit \cite{Rosado}. Therefore it can be characterized by the fluorescence yield, $Y_{\lambda}$, defined as the number of photons per unit of deposited energy at each N$_2$ emission band centered at wavelength $\lambda$. The fluorescence yield depends on the air pressure, temperature and humidity due to the process of collisional quenching. A full description of the atmospheric fluorescence produced by EASs can be given by the following expression (see, e.g., \cite{AIRFLY1}):

\begin{equation}
\label{parametrization}
Y_{\lambda}(P,T,h)=Y_{337}(P_0,T_0,h_0)\,I_{\lambda}(P_0,T_0,h_0)
\frac{1+\frac{P_0}{P'_{337}(T_0,h_0)}}{1+\frac{P}{P'_{\lambda}(T,h)}}\,,
\end{equation}
where $Y_{337}(P_0,T_0,h_0)$ is the absolute value of the fluorescence yield for the reference molecular band at 337~nm (i.e., the most intense one) at a certain pressure, temperature and humidity conditions, $I_{\lambda}(P_0,T_0,h_0)$ are the intensities of the other bands relative to this reference one and at the same air conditions, and $P'_{\lambda}(T,h)$ is a quenching parameter that represents the air pressure at which the collisional de-excitation rate equals the radiative one. This parameter depends on temperature and humidity.

In this work, we have used Eqn. (\ref{parametrization}) with the world average $Y_{337}(P_0,T_0,h_0)$ reported in \cite{Rosado} and the other parameters were taken from \cite{AIRFLY1,AIRFLY2}. The humidity dependence was however ignored since the standard atmospheric parameterization in CORSIKA has no information on water vapor concentration. The air temperature was estimated under the assumption that air is an ideal gas. For the tests presented here, we employed the U.S. standard atmosphere. Fluorescence emission was assumed to be instantaneous and no time spread from this process was included for the tests performed.

\subsection{Numerical model}
\label{ssec:Calculations}

For each of the simulated EAS configurations, numerical calculations were performed using a simplified model that starts from the longitudinal energy deposit calculated by CORSIKA. The atmospheric model and the fluorescence parameterization were identical to the ones used in the simulations. The same shower-detector geometry of the simulations was reproduced in the calculations too, where the detector area was sampled on a grid of {\it observation points} on the ground.

The atmosphere was discretized in horizontal layers of 200~m width. For each shower element delimited within one layer, the flux of fluorescence and Cherenkov photons at every observation point was calculated. In the case of fluorescence, the total light intensity is proportional to the energy deposited in the given shower element and geometrical factors are easy to calculate due to the fact that fluorescence is emitted isotropically. For Cherenkov radiation, we used the parameterizations
given in \cite{Nerling}, which allow the calculation of the flux as a function of the energy deposit profile based on the universality of both the energy and angular distributions of electrons in the EASs.

The total flux of one or the other component reaching a given observation point was obtained summing the contributions from all the shower elements. Alternatively, the summation was restricted to shower elements that are within a certain field of view from the observation point.

\section{Results}
\label{sec:Results}

The above-described tools have been used to quantify the fluorescence contamination of Che\-ren\-kov light detectors signals in typical working conditions. 
For this work, we defined a rectangular and continuous detector of $10 \times 10$ km$^2$ size, centered at the intersection of the shower axis with the ground, which is set to be at 2200~m above sea level. The light density at the ground has been obtained separately for Cherenkov and fluorescence photons as a function of the core distance. Light scattering and absorption were not included.

\begin{figure}[H]
\centering
\includegraphics[width=0.495\textwidth]{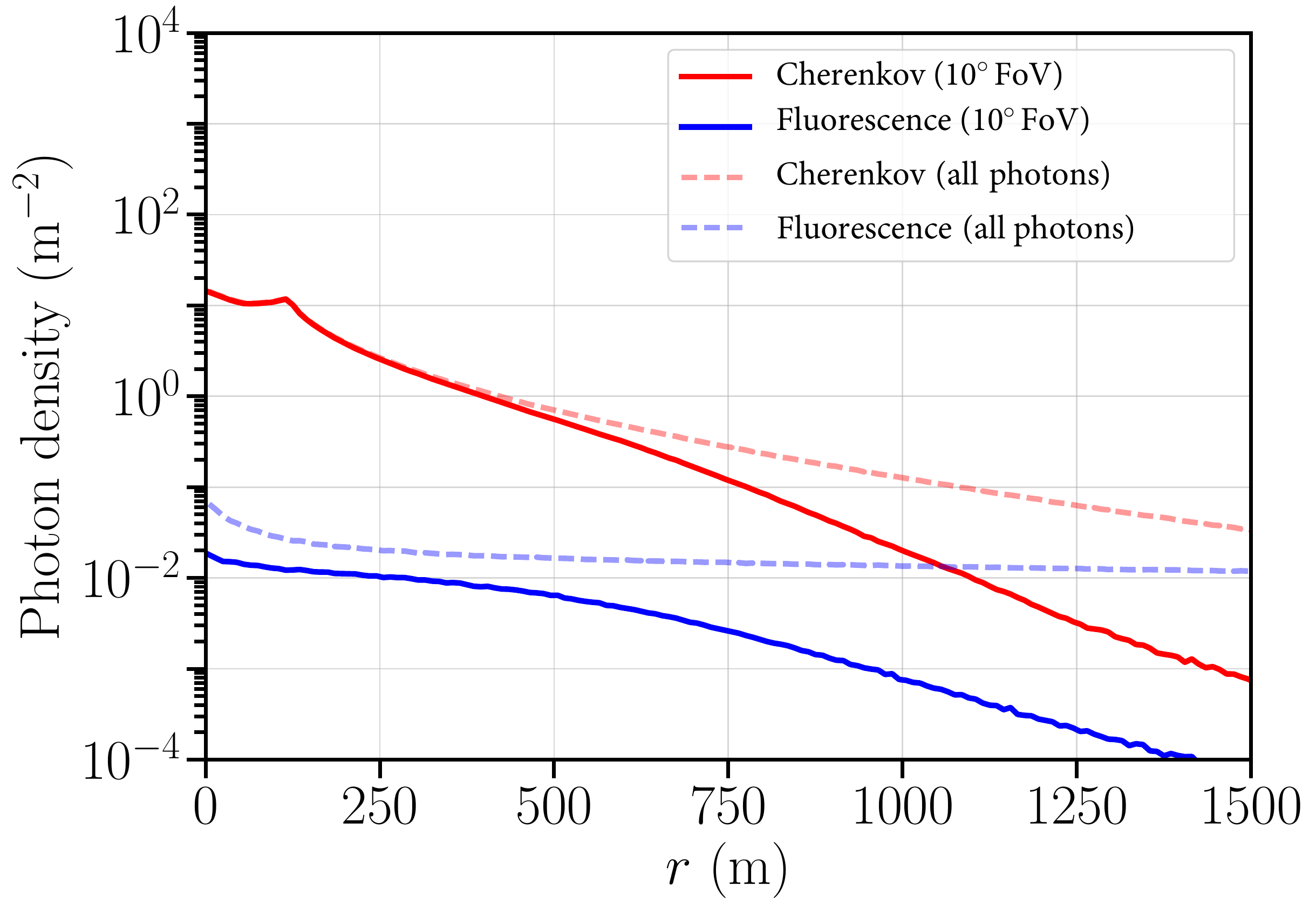} 
\includegraphics[width=0.495\textwidth]{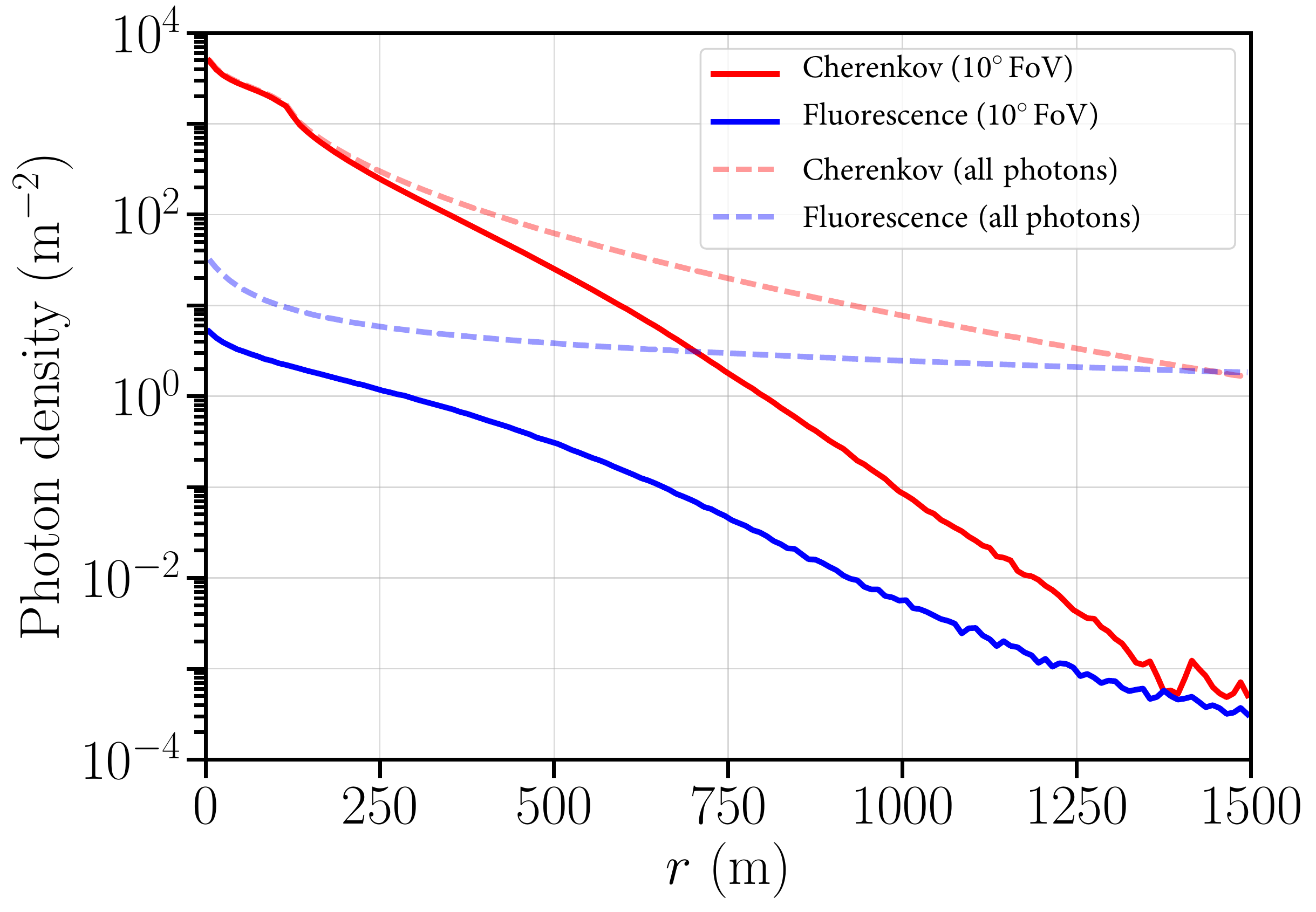}
\caption{Photon density versus radial distance for vertical showers including or not including field of view constraints. \textit{Left:} 100 GeV $\gamma$-rays. \textit{Right:} 10 TeV $\gamma$-rays.}
\label{fig:radial_hist}
\end{figure}

Figure~\ref{fig:radial_hist} displays the average simulation results of a number of vertical showers initiated by 100~GeV (left) and 10~TeV $\gamma$-rays (right). Dotted lines correspond to total photon densities while solid ones represent those arriving within an angular radius of $5^\circ$ around the vertical direction, that is, the case of a Cherenkov telescope with a field of view of $10^\circ$ pointing to a vertical source. The angular constraint reduces significantly the photon flux at radial distances of some hundreds of meters, because the farther a telescope is, the shorter and higher is the section of the shower it observes.

The main feature that can be extracted from these plots is that the fluorescence contamination in Cherenkov telescopes increases steeply at increasing impact parameters, i.e., from $\sim 0.1\%$ very near the shower axis to $\sim 1\%$ at some hundreds of meters. This is a consequence of the much flatter radial distribution of fluorescence radiation arising from its isotropic nature. 

\begin{figure}[H]
\centering
\includegraphics[width=0.495\textwidth]{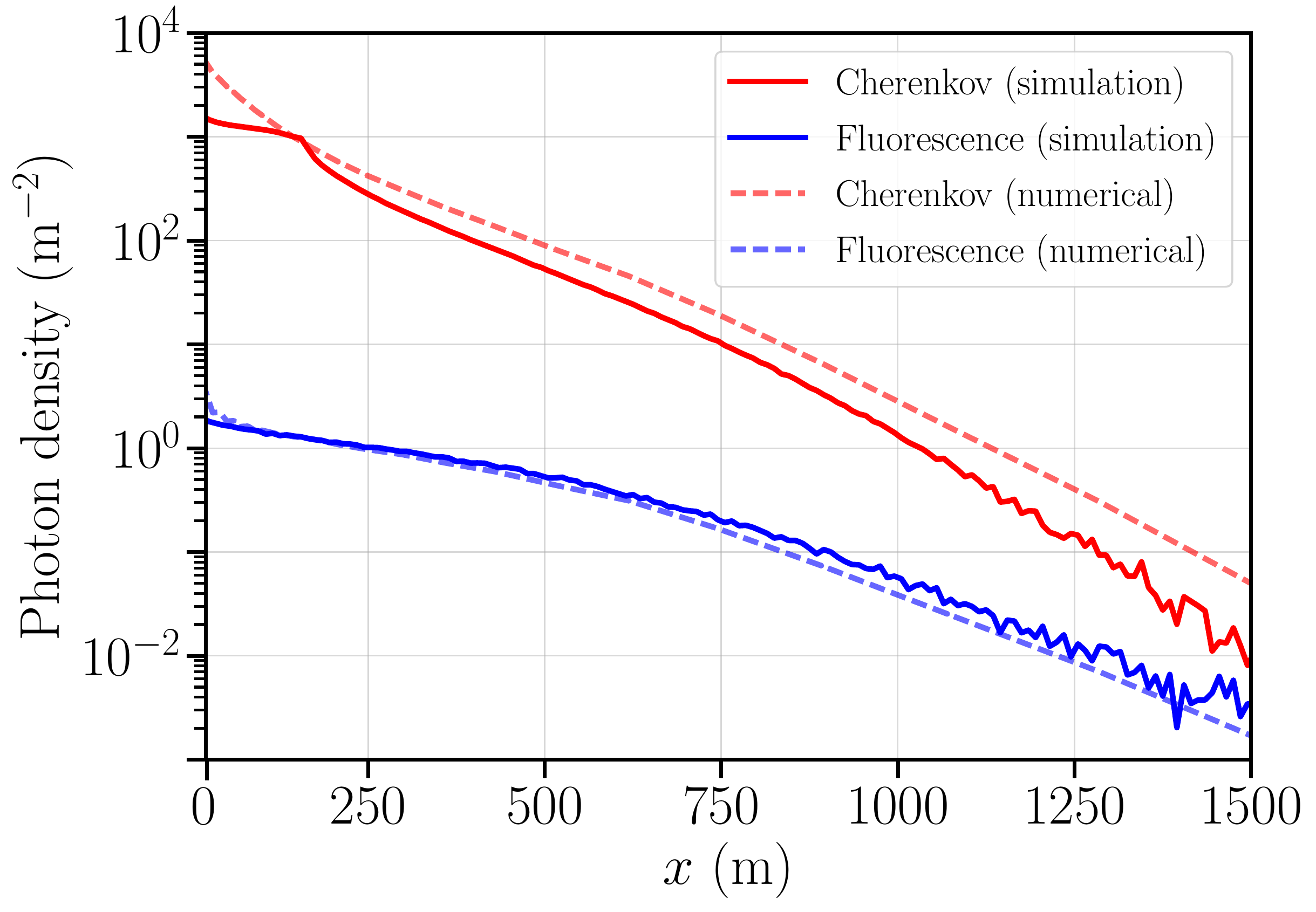}
\includegraphics[width=0.495\textwidth]{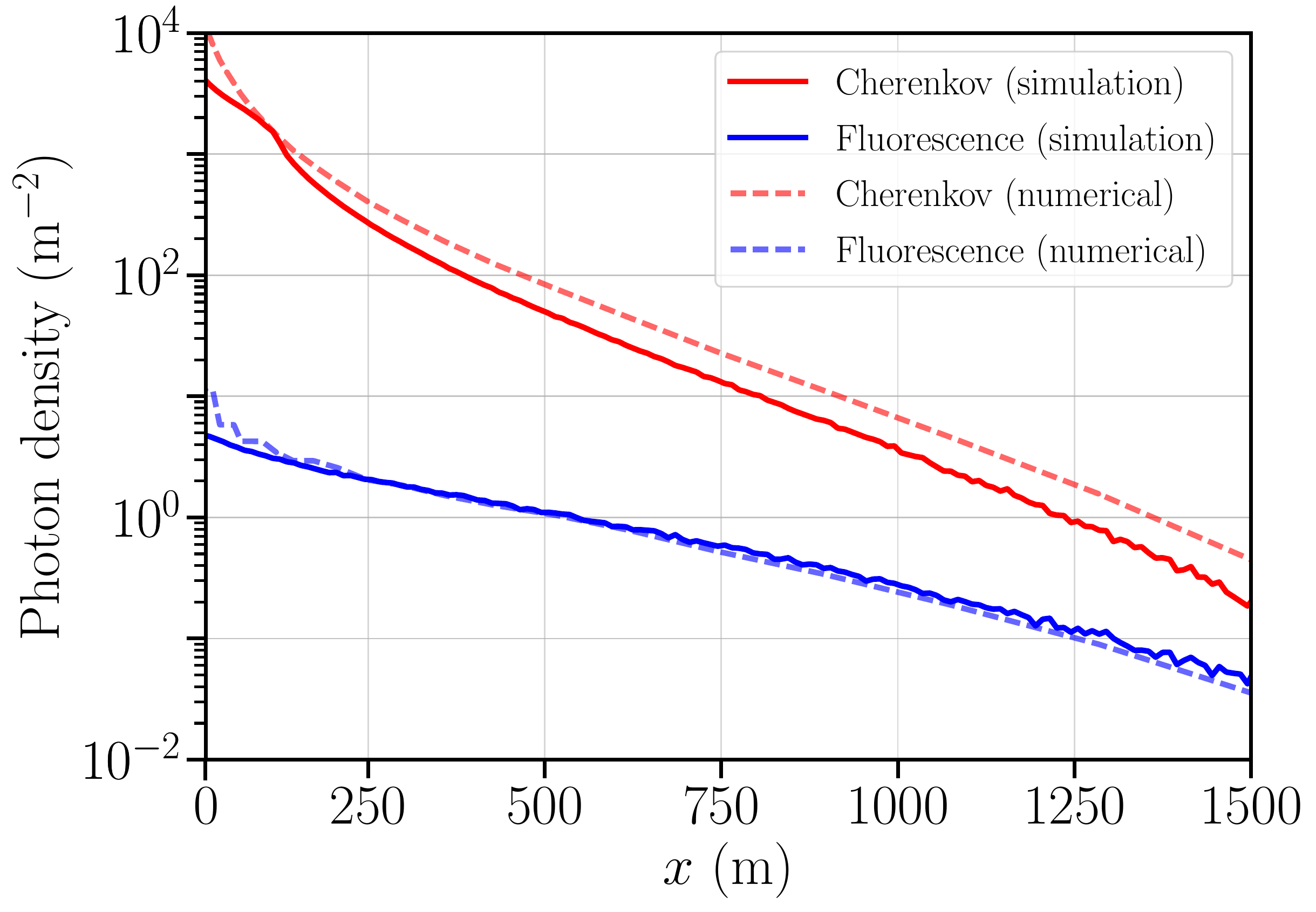}
\caption{Photon density versus distance along the ground projection of the shower axis for two different configurations. \textit{Left:} 10~TeV shower with zenith angle of $30^\circ$ observed on-axis. \textit{Right:} 10~TeV nearly vertical shower observed with an off-axis angle of $5^\circ$. Simulation results are compared with those obtained by the numerical model.}
\label{fig:other_config}
\end{figure}

Tests have also been performed for observations at large zenith angle for which the EAS has to penetrate a larger atmospheric thickness. For example, the left-hand plot of figure~\ref{fig:other_config} shows the results for 10~TeV showers with a zenith angle $\theta=30^\circ$. The photon density distributions were only evaluated along the ground projection of the shower axis $x$. The fluorescence contamination is somewhat lower than for vertical showers (e.g., $1.0\%$ instead of $1.2\%$ at 500~m when comparing with the right-hand plot of figure~\ref{fig:radial_hist}), because inclined showers developed farther away from the telescope.

The finite field of view of Cherenkov telescopes leads to many EASs being observed with a certain off-axis angle $\Delta\theta$. The effect of an offset of $5^\circ$ on 10~TeV nearly vertical showers is presented in the right-hand plot of figure~\ref{fig:other_config}. The comparison with the on-axis case (right-hand plot of figure~\ref{fig:radial_hist}) indicates that the fluorescence contamination is increased by the offset (from $1.2\%$ up to $2.3\%$ at 500~m) because the field of view of the telescopes covers a larger and closer segment of the EAS, favoring the fluorescence component.

The simulations have been compared with numerical calculations obtained from the average longitudinal profile of energy deposit of the simulated showers (see section~\ref{ssec:Calculations}). As an example, the comparison is shown for the two cases of 10~TeV showers of figure~\ref{fig:other_config}. The numerical results of fluorescence density are in good agreement with the simulations in the whole range of distances. However the numerical model overestimates the Cherenkov component by a factor of about 2 and does not reproduce correctly the {\it Cherenkov hump}, very likely due to the analytical approximations of the angular distributions of shower electrons. In spite of this disagreement, the fluorescence contamination can be safely estimated for other cases using our numerical method. For instance, we have found that it is basically the same in the whole energy range from 10~GeV to 1~PeV at a given core distance, although there exists a very small increase with energy because higher-energy EASs develop closer to the ground, increasing the density of fluorescence light at ground level to a greater extent than that of Cherenkov light, which is strongly directional.

From the results shown here, we can conclude that the fluorescence contamination in Cherenkov telescopes signals is around $1\%$ at impact parameters of about 500~m. For high energy showers, this contamination can be significant in arrays of telescopes that may collect measurable signals at large core distances. Observations of low energy showers are presently restricted to telescopes inside the Cherenkov light pool and therefore the fluorescence contamination is of the order of $0.1\%$ according to our results. 

\section{Further considerations}
\label{sec:Further}

Since the evaluation of the fluorescence emission in EAS simulations might be of general interest, we propose to include this implementation as an option in future versions of CORSIKA.

In \cite{ICRC2015} a simple model was used to study the capability of IACT as fluorescence telescopes. In this technique, the telescope should not point to the source but observe the shower laterally. Since the field of view of a Cherenkov telescope is rather small, the telescope only records a small fraction of the shower track. However, when using an array of IACTs working stereoscopically, the information could be enough to measure accurately the arrival direction. 
In addition, the excellent angular resolution of IACTs would allow a detailed sampling of the transversal distribution of the energy deposit, which carries valuable information on the nature of the primary particle. Transit time in fluorescence mode will be in general larger than the typical time window of Cherenkov telescopes and therefore some modifications would be needed for this application.
The procedure presented in this work will be used to extend the study initiated in \cite{ICRC2015}.

\section{Acknowledgments}
This work was supported by the Spanish MINECO under contract 
FPA2015-69210-C6-3-R.

\end{document}